# Bulk-edge correspondence for topological photonic continua

*Mário G. Silveirinha*[*]

[1]*University of Lisbon–Instituto Superior Técnico and Instituto de Telecomunicações, Avenida Rovisco Pais, 1, 1049-001 Lisboa, Portugal*

**Abstract**

Here, building on our previous work [*Phys. Rev. B*, **92**, 125153, (2015)], it is shown that the propagation of unidirectional gapless edge states at an interface of two topologically distinct electromagnetic continua with a well-behaved asymptotic electromagnetic response is rigorously predicted by the bulk-edge correspondence principle. We work out detailed examples demonstrating that when the spatial-cut off of the nonreciprocal part of the material response is considered self-consistently in the solution of the relevant electromagnetic problem, the number of unidirectional gapless edge modes is identical to the difference of the Chern numbers of the bulk materials. Furthermore, it is shown how the role of the spatial cut-off can be imitated in realistic systems using a tiny air gap with a specific thickness. This theory provides a practical roadmap for the application of topological concepts to photonic platforms formed by nonreciprocal electromagnetic continua.

**PACS numbers:** 42.70.Qs, 03.65.Vf, 78.67.Pt, 78.20.Ls

---

[*] To whom correspondence should be addressed: E-mail: *mario.silveirinha@co.it.pt*

# I. Introduction

Topological materials have emerged in the last decade as one of the most exciting research topics both in electronics and in photonics [1-4]. The intimate connection between the abstract topological properties of the band structure and the wave propagation with no back-scattering made possible the development of novel paradigms and platforms for waveguiding insensitive to disorder and imperfections [5-13]. In particular, the influential works by Raghu and Haldane established that nonreciprocal (e.g., gyromagnetic) photonic crystals can have a nontrivial topology and may be characterized by a nonzero topological (Chern) invariant [5-6]. When a topological material is paired with another material with a trivial topology, unidirectional gapless edge states appear in a common bandgap [3, 9, 12].

A nontrivial Chern number requires a broken time-reversal symmetry (nonreciprocal response), which is typically obtained with a biasing magnetic field. There has a been a significant effort in the recent literature to extend the use of topological concepts to standard reciprocal materials (e.g. dielectrics and metals), exploring for example a bianisotropic response [14-15], using non-periodic optical potentials to mimic a synthetic pseudo-magnetic field [16-18], taking advantage of particular symmetries [19, 20], or by imitating the role of time with a specific spatial coordinate (Floquet topological insulators) [21, 22].

The characterization of topological phases of a material generally requires that the underlying structure is spatially periodic so that the wave vector space (the Brillouin zone) is a closed manifold with no boundary. Interestingly, a few recent works have shown that topological methods may be extended to an electromagnetic continuum with no intrinsic periodicity. One approach, in the context of Floquet topological insulators, relies on the computation of the Chern invariants of equifrequency surfaces, the electromagnetic analogue of the Fermi surface [22]. Related ideas have also been explored in Ref. [23].

A totally different approach was introduced by us in Ref. [24], and the key idea is that there is a class of sufficiently well-behaved (bianisotropic) material responses for which the Chern number of the Euclidean plane (an unbounded open manifold) is guaranteed to be an integer. Moreover, for the relevant electromagnetic continua the Euclidean plane can be identified with the Riemann sphere [24]. The pertinent material class is formed by media such that the nonreciprocal part of the electromagnetic response is negligible in the limit $\mathbf{k} \to \infty$, whereas for finite $\mathbf{k}$ the response in unconstrained. It was explained how the response of standard materials can be easily modified to satisfy this requirement by introducing a spatial cut-off in the nonreciprocal part of the material matrix [24]. In particular, the response of any given conventional material can be arbitrarily well approximated (in any compact subset of the spectral $\mathbf{k}$-plane) by that of an element in the class of electromagnetic continua with integer Chern numbers. The proposed framework enables one to extend standard topological concepts, such as the Berry phase, the Berry curvature, Chern numbers, to dispersive electromagnetic continua, and can be used to characterize topological phases of nonreciprocal lossless materials [24, 25, 26]. Furthermore, in Ref. [27] we also demonstrated that – analogous to electronic systems (electronic topological insulators) – dispersive electromagnetic continua with the time reversal symmetry can be separated into two topological phases. Specifically, a dispersive reciprocal material is characterized by a $\mathbb{Z}_2$ topological invariant, which is nonzero when there is an obstruction to the application of the Stokes theorem to half-wave vector space.

Perhaps the most celebrated property of topological systems is the *bulk-edge correspondence* that links the Chern invariants of two inequivalent topological materials with the number of gapless unidirectional edge states propagating at a material interface in a common frequency bandgap. As previously explained, even though typical nonreciprocal responses do not have the desired asymptotic behavior in the $\mathbf{k} \to \infty$ limit, they can be arbitrarily well approximated by an element in the class of well-behaved media with integer Chern numbers. A bit surprisingly, it was verified in Ref. [24] that the bulk-edge correspondence does not always hold in the continuum limit when the spatial cut-off is disregarded in the calculation of the edge modes. Specifically, the Chern numbers of the bulk materials are computed by considering "well-behaved" materials with a spatial cut-off in the nonreciprocal response, and that approximate arbitrarily well the original materials. Crucially, it was verified in Ref. [24] that the edge modes supported by an interface of the original materials (with no high-frequency spatial cut-off) do not satisfy the bulk-edge correspondence principle. Indeed, the number of modes may be inconsistent with the bulk-edge correspondence and moreover the edge modes dispersion may not span the entire bandgap.

The objective of the present article is twofold. The first goal is to demonstrate that the bulk-edge correspondence is rigorously satisfied when the spatial-cut off is considered self-consistently, i.e., both in the calculation of the Chern numbers and in the calculation of edge modes. This result is important mainly from a theoretical standpoint, because the usual material responses of realistic media do not have an *explicit* spatial cut-off. The second objective is to demonstrate, building on the ideas of Ref. [24], that the role of spatial cut-off can be imitated in conventional systems simply by inserting a tiny air gap in between the relevant materials. It is shown that the thickness of the air gap is inversely proportional to the spatial cut-off and that this solution is particularly effective for edge modes with propagating wave numbers comparable to the spatial cut-off. It is numerically shown with detailed examples that the modified material interface always supports a number of unidirectional edge modes consistent with what is predicted by the bulk-edge correspondence principle, even for high-order Chern number differences ($|\delta \mathcal{C}| > 1$).

## II. Topological methods in a photonic continuum

Here, we briefly review the topological concepts introduced in Ref. [24]. Let us consider a homogeneous photonic continuum described by a material matrix $\mathbf{M}$. In a continuum the stationary states of the Maxwell's equations are plane waves and can be labeled by the wave vector $\mathbf{k}$. The envelope of a plane wave $\mathbf{f}_{n\mathbf{k}} = \begin{pmatrix} \mathbf{E}_{n\mathbf{k}} & \mathbf{H}_{n\mathbf{k}} \end{pmatrix}^T$ satisfies:

$$\left[ \hat{N}(\mathbf{k}) - \omega \mathbf{M}(\omega, \mathbf{k}) \right] \cdot \mathbf{f}_{n\mathbf{k}} = 0, \tag{1}$$

where $\hat{N} = \begin{pmatrix} \mathbf{0} & -\mathbf{k} \times \mathbf{1}_{3\times 3} \\ \mathbf{k} \times \mathbf{1}_{3\times 3} & \mathbf{0} \end{pmatrix}$. Note that $\mathbf{f}_{n\mathbf{k}}$ is a six-component vector formed by the electric ($\mathbf{E}_\mathbf{k}$) and the magnetic ($\mathbf{H}_\mathbf{k}$) fields envelopes. The symbol "$T$" denotes the transpose operator and the subscript "$n$" identifies a particular family of eigenmodes. In this article, it is assumed that the material matrix is of the form $\mathbf{M} = \begin{pmatrix} \varepsilon_0 \boldsymbol{\varepsilon} & \mathbf{0} \\ \mathbf{0} & \mu_0 \mathbf{1} \end{pmatrix}$ so that the material response is purely electric. In a photonic continuum the Berry potential is given by [5, 6, 24]:

$$\mathcal{A}_{n\mathbf{k}} = \frac{\text{Re}\left\{i\mathbf{f}_{n\mathbf{k}}^* \cdot \frac{\partial}{\partial \omega}\left[\omega \mathbf{M}(\omega,\mathbf{k})\right]_{\omega_{n\mathbf{k}}} \cdot \partial_\mathbf{k} \mathbf{f}_{n\mathbf{k}}\right\}}{\mathbf{f}_{n\mathbf{k}}^* \cdot \frac{\partial}{\partial \omega}\left[\omega \mathbf{M}(\omega,\mathbf{k})\right]_{\omega_{n\mathbf{k}}} \cdot \mathbf{f}_{n\mathbf{k}}}. \tag{2}$$

The Chern number associated with a subset of photonic bands is given by

$$\mathcal{C} = \frac{1}{2\pi} \iint dk_x dk_y \, \mathcal{F}_\mathbf{k}, \tag{3}$$

where $\mathcal{F}_\mathbf{k} = \frac{\partial \mathcal{A}_y}{\partial k_x} - \frac{\partial \mathcal{A}_x}{\partial k_y}$ is the Berry curvature. In a continuum, the wave vector space is the Euclidian plane (e.g., the plane $k_z = 0$) including the point $k = \infty$. As discussed in Ref. [24], it is often useful to regard the wave vector space as the Riemann sphere. For photonic continua invariant under arbitrary rotations about the $z$-axis the Chern number can be calculated using $\mathcal{C} = \lim_{k \to \infty}\left(\mathcal{A}_{\varphi=0} k\right) - \lim_{k \to 0^+}\left(\mathcal{A}_{\varphi=0} k\right)$ [24], where $(k, \varphi)$ determines a system of polar coordinates in the $\mathbf{k}$-plane and $\mathcal{A}_{n,\varphi} = \mathcal{A}_{n\mathbf{k}} \cdot \hat{\varphi}$.

It was proven in Ref. [24] that for general material responses of the form:

$$\boldsymbol{\varepsilon}(\omega, \mathbf{k}) = \boldsymbol{\varepsilon}_R(\omega) + \frac{1}{1 + k^2/k_{\max}^2} \boldsymbol{\chi}_{NR}(\omega), \tag{4}$$

the Chern numbers are integer. Here, $\boldsymbol{\varepsilon}_R(\omega)$ is a symmetric matrix $\boldsymbol{\varepsilon}_R(\omega) = \left[\boldsymbol{\varepsilon}_R(\omega)\right]^T$, so that it determines a reciprocal (time-reversal invariant) response. In general, the susceptibility $\boldsymbol{\chi}_{NR}(\omega)$ gives the nonreciprocal component of the material response that breaks the time-reversal symmetry. In order that the Chern number is an integer this component is required to be associated with a spatial cut-off determined by $k_{\max}$ so that for large values of $k$ ($k^2 = \mathbf{k} \cdot \mathbf{k}$) the response becomes reciprocal. Hence, to compute the Chern numbers associated with some photonic continuum in general one needs to impose by hand a spatial cut-off in the material response [24].

## III. Edge modes

In order to study the validity of the bulk-edge correspondence, in this section we obtain the dispersion equation for the edge states supported by a planar interface that separates two topologically different photonic continua. It is assumed that the dielectric function of the relevant materials is as in Eq. (4), and hence the materials are spatially dispersive (nonlocal) because the material matrix depends explicitly on the wave vector [28, 29].

### A.  *Space domain formulation*

As a first step in the calculation the edge modes, next we obtain a space domain formulation of the electromagnetic problem. Specifically, it is shown that by introducing an auxiliary field it is possible to get rid of the spatial dispersion in the material response. Indeed, let us define the auxiliary field $\tilde{\mathbf{E}} = \frac{1}{1 + k^2 / k_{\max}^2} \boldsymbol{\chi}_{\mathrm{NR}} \cdot \mathbf{E}$. Then, Ampère's law in differential form, $\nabla \times \mathbf{H} = -i\omega\varepsilon_0 \boldsymbol{\varepsilon} \cdot \mathbf{E}$, can be conveniently rewritten in the spatial domain as:

$$\nabla \times \mathbf{H} = -i\omega\varepsilon_0 \boldsymbol{\varepsilon}_{\mathrm{R}}(\mathbf{r}) \cdot \mathbf{E} - i\omega\varepsilon_0 \tilde{\mathbf{E}}, \tag{5a}$$

$$\left( -\frac{1}{k_{\max}^2} \nabla^2 + 1 \right) \tilde{\mathbf{E}} = \boldsymbol{\chi}_{\mathrm{NR}}(\mathbf{r}) \cdot \mathbf{E}, \tag{5b}$$

The electrodynamics of the original spatially dispersive material [Eq. (4)] is precisely described by Eqs. (5a)-(5b) complemented with Faraday's law:

$$\nabla \times \mathbf{E} = i\omega\mu_0 \mathbf{H}. \qquad (5c)$$

In particular, for a bulk material the plane wave natural modes determined by the system (5) are exactly the same as the modes directly found from the nonlocal dielectric function (4). Importantly, the system (5) is formulated in the space domain, and hence one may admit that $\boldsymbol{\chi}_{NR}$ and $\boldsymbol{\varepsilon}_R$ are space dependent, as explicitly indicated in the formulas. Thus, the system (5) is suitable to determine the impact of the nonlocal effects on the edge modes.

### B. *TM-polarized plane waves*

We are interested in transverse magnetic (TM) polarized waves with $\partial/\partial z = 0$ (*xoy* plane propagation), whose nontrivial electromagnetic field components are $E_x, E_y, H_z$. It is supposed that the nonlocal dielectric function is of the form:

$$\boldsymbol{\varepsilon}(\omega,\mathbf{k}) = \begin{pmatrix} \varepsilon_{11} & \varepsilon_{12} & 0 \\ \varepsilon_{21} & \varepsilon_{22} & 0 \\ 0 & 0 & \varepsilon_{33} \end{pmatrix}, \qquad (6)$$

with $\varepsilon_{11} = \varepsilon_{22}$ and $\varepsilon_{12} = -\varepsilon_{21}$. Similarly, the tensor $\boldsymbol{\chi}_{NR}(\omega)$ in Eq. (4) is assumed of the form:

$$\boldsymbol{\chi}_{NR} = \begin{pmatrix} \chi_{11} & \chi_{12} & 0 \\ \chi_{21} & \chi_{22} & 0 \\ 0 & 0 & 0 \end{pmatrix}, \qquad (7)$$

with $\chi_{11} = \chi_{22}$ and $\chi_{12} = -\chi_{21}$. In section IV we will discuss examples of materials with such a response. For a plane-wave type mode with a spatial variation $e^{i\mathbf{k}\cdot\mathbf{r}}$ ($\partial_x \equiv \partial/\partial x = ik_x$ and $\partial_y \equiv \partial/\partial y = ik_y$), the electric field ($\mathbf{E} = E_x\hat{\mathbf{x}} + E_y\hat{\mathbf{y}}$) and the auxiliary field ($\tilde{\mathbf{E}} = \tilde{E}_x \hat{\mathbf{x}} + \tilde{E}_y \hat{\mathbf{y}}$) can be written in terms of the magnetic field ($\mathbf{H} = H_z \hat{\mathbf{z}}$) as:

$$E_x = \theta_1 \partial_y H_z + \theta_2 \partial_x H_z, \qquad E_y = \theta_2 \partial_y H_z - \theta_1 \partial_x H_z, \qquad (8a)$$

$$\tilde{E}_x = \Theta_1 \partial_y H_z + \Theta_2 \partial_x H_z, \qquad \tilde{E}_y = \Theta_2 \partial_y H_z - \Theta_1 \partial_x H_z. \qquad (8b)$$

The $\theta_{1,2}$ and $\Theta_{1,2}$ coefficients are defined as,

$$\theta_1 = \frac{1}{-i\omega\varepsilon_0} \frac{\varepsilon_{11}}{\varepsilon_{11}^2 + \varepsilon_{12}^2}, \qquad \theta_2 = \frac{1}{-i\omega\varepsilon_0} \frac{\varepsilon_{12}}{\varepsilon_{11}^2 + \varepsilon_{12}^2}, \qquad (9a)$$

$$\Theta_1 = \frac{1}{-i\omega\varepsilon_0} \frac{1}{1 + k^2/k_{max}^2} \frac{1}{\varepsilon_{11}^2 + \varepsilon_{12}^2} (\chi_{11}\varepsilon_{11} + \chi_{12}\varepsilon_{12}),$$

$$\Theta_2 = \frac{1}{-i\omega\varepsilon_0} \frac{1}{1 + k^2/k_{max}^2} \frac{1}{\varepsilon_{11}^2 + \varepsilon_{12}^2} (\chi_{11}\varepsilon_{12} - \chi_{12}\varepsilon_{11}). \qquad (9b)$$

The TM-polarized plane wave modes satisfy the standard dispersion equation [24]:

$$k^2 = \frac{\varepsilon_{11}^2 + \varepsilon_{12}^2}{\varepsilon_{11}} \left(\frac{\omega}{c}\right)^2, \qquad \text{(TM modes)}. \qquad (10)$$

It turns out that because of the nonlocal effects ($\boldsymbol{\varepsilon} = \boldsymbol{\varepsilon}(\omega, \mathbf{k})$) Eq. (10) has three different solutions for a fixed direction of space, or in other words there are three branches of TM-waves. Thus, as compared to the local formulation ($k_{max} = \infty$) for which there is a single TM branch, the spatial dispersion leads to the appearance of two new modes. The emergence of additional waves is a well known property of spatially dispersive media [28, 32-40].

## C. *Edge modes at a planar interface of two nonlocal materials*

We want to determine the dispersion of the edge modes at an interface (plane $y = 0$) between two different materials with a dielectric response as in Eq. (4). To this end, we look for solutions of the system (5) [which is formulated in the spatial domain] with a magnetic field of the form $H_z = h(y)e^{ik_x x}$, where $k_x$ is the propagation constant of the edge mode along the $x$-direction. In each homogenous semi-space the magnetic field can be written as superposition of decaying (along the direction normal to the interface) exponentials as follows:

$$H_z = e^{ik_x x} \begin{cases} A_1 e^{-\gamma_{a1} y} + A_2 e^{-\gamma_{a2} y} + A_3 e^{-\gamma_{a3} y}, & y > 0 \\ -B_1 e^{+\gamma_{b1} y} - B_2 e^{+\gamma_{b2} y} - B_3 e^{+\gamma_{b3} y}, & y < 0 \end{cases}, \quad (11)$$

where $A_i, B_i$ ($i = 1, 2, 3$) are unknown constants. In each region the magnetic field is a superposition of three plane wave modes, consistent with the fact that Eq. (10) gives three distinct solutions $k_{s,1}^2$, $k_{s,2}^2$ and $k_{s,3}^2$, when solved with respect to $k^2$. The attenuation constant along the $y$-direction is $\gamma_{s,i} = \sqrt{k_x^2 - k_{s,i}^2}$, being $s = a$ ($s = b$) in the region $y > 0$ ($y < 0$), respectively.

As usual, the electromagnetic fields need to satisfy suitable boundary conditions at the interfaces. In particular, one should impose the continuity of $H_z, E_x$ (the standard Maxwellian boundary conditions) at the interface $y = 0$. In addition, to guarantee that the right-hand side of Eq. (5b) has no delta-functions, i.e., that the electric field is piecewise continuous, one also needs to enforce that $\tilde{E}_x$, $\tilde{E}_y$ and $\partial_y \tilde{E}_x$, $\partial_y \tilde{E}_y$ are continuous at the interface. Thus, as compared to a local formulation, one needs to impose *four* additional boundary conditions (ABCs). These boundary conditions emerge naturally from the system (5), and are consistent with the fact that each bulk medium supports two additional waves. In fact, the total number of additional waves (2+2=4) is equal to the number of ABCs, as it should be. Additional boundary conditions have been extensively discussed in the literature in the context of electromagnetic metamaterials [32-37] and of plasmonics [38-40].

Taking into account that the electric field and the auxiliary field can be written in terms of $H_z$ using Eq. (8) and imposing the six boundary conditions, it is straightforward to show that the unknown coefficients must satisfy:

$$\sum_i A_i + \sum_i B_i = 0, \tag{12a}$$

$$\sum_i \left(-\theta_{1,i}^a \gamma_{a,i} + \theta_{2,i}^a ik_x\right) A_i + \sum_i \left(+\theta_{1,i}^b \gamma_{b,i} + \theta_{2,i}^b ik_x\right) B_i = 0, \tag{12b}$$

$$\sum_i \left(-\Theta_{1,i}^a \gamma_{a,i} + \Theta_{2,i}^a ik_x\right) A_i + \sum_i \left(+\Theta_{1,i}^b \gamma_{b,i} + \Theta_{2,i}^b ik_x\right) B_i = 0, \tag{12c}$$

$$\sum_i \left(-\Theta_{2,i}^a \gamma_{a,i} - \Theta_{1,i}^a ik_x\right) A_i + \sum_i \left(+\Theta_{2,i}^b \gamma_{b,i} - \Theta_{1,i}^b ik_x\right) B_i = 0, \tag{12d}$$

$$\sum_i \left(\Theta_{1,i}^a \gamma_{a,i}^2 - \gamma_{a,i} \Theta_{2,i}^a ik_x\right) A_i + \sum_i \left(\Theta_{1,i}^b \gamma_{b,i}^2 + \gamma_{b,i} \Theta_{2,i}^b ik_x\right) B_i = 0, \tag{12e}$$

$$\sum_i \left(\Theta_{2,i}^a \gamma_{a,i}^2 + \gamma_{a,i} \Theta_{1,i}^a ik_x\right) A_i + \sum_i \left(\Theta_{2,i}^b \gamma_{b,i}^2 - \gamma_{b,i} \Theta_{1,i}^b ik_x\right) B_i = 0. \tag{12f}$$

In the above, $\theta_{1,i}^s$, $\theta_{2,i}^s$, $\Theta_{1,i}^s$ and $\Theta_{2,i}^s$ stand for $\theta_1$, $\theta_2$, $\Theta_1$ and $\Theta_2$ [defined as in Eq. (9)] evaluated with $k^2 = k_{s,i}^2$ ($i = 1, 2, 3$) and $s = a, b$ depending on the considered region of space. The dispersion of the edge modes is obtained by setting the determinant of the matrix associated with the 6×6 homogeneous linear system (12) equal to zero.

For future reference, it is noted that in the local case ($k_{max} = \infty$) the dispersion of the edge modes of the same two materials, eventually separated by an air gap with thickness $d$, is given by (for conciseness the derivation is omitted):

$$\left(\frac{\gamma_a}{\varepsilon_{ef,a}} - \frac{\varepsilon_{12,a}ik_x}{\varepsilon_{11,a}^2 + \varepsilon_{12,a}^2}\right) + \left(\frac{+\gamma_b}{\varepsilon_{ef,b}} + \frac{\varepsilon_{12,b}ik_x}{\varepsilon_{11,b}^2 + \varepsilon_{12,b}^2}\right) + \frac{\gamma_{air}}{\varepsilon_{air}}\tanh(\gamma_{air}d) +$$
$$+ \left(\frac{\gamma_a}{\varepsilon_{ef,a}} - \frac{\varepsilon_{12,a}ik_x}{\varepsilon_{11,a}^2 + \varepsilon_{12,a}^2}\right)\left(\frac{+\gamma_b}{\varepsilon_{ef,b}} + \frac{\varepsilon_{12,b}ik_x}{\varepsilon_{11,b}^2 + \varepsilon_{12,b}^2}\right)\frac{\varepsilon_{air}}{\gamma_{air}}\tanh(\gamma_{air}d) = 0 \quad (13)$$

The index $s = a$ ($s = b$) refers to the material region $y > d$ ($y < 0$). The transverse attenuation constant $\gamma_s = \sqrt{k_x^2 - k_s^2}$ ($s = a,b$) is defined as in the local case, but now $k_s^2 = \varepsilon_{ef,s}(\omega/c)^2$ is the unique solution of Eq. (10) with respect to $k^2$. We use the notation $\varepsilon_{ef,s} = (\varepsilon_{11,s}^2 + \varepsilon_{12,s}^2)/\varepsilon_{11,s}$. Moreover, we define the transverse attenuation constant in the air region as $\gamma_{air} = \sqrt{k_x^2 - \varepsilon_{air}(\omega/c)^2}$, being $\varepsilon_{air} = 1$ the relative permittivity of the air gap. Evidently, when the air gap has vanishing thickness the dispersion equation is determined only by the first two terms in brackets in the left-hand side of Eq. (13).

## IV. Bulk-Edge Correspondence

To illustrate the application of the developed theory we consider a nonreciprocal gyrotropic bulk material characterized by a permittivity with a spatial cut-off:

$$\varepsilon_{11} = \varepsilon_{22} = 1 + \frac{\omega_0\omega_e}{\omega_0^2 - \omega^2}\frac{1}{1 + k^2/k_{max}^2}, \qquad \varepsilon_{12} = -\varepsilon_{21} = \frac{-i\omega_e\omega}{\omega_0^2 - \omega^2}\frac{1}{1 + k^2/k_{max}^2}. \quad (14)$$

For example, a magnetized electric plasma has a similar material dispersion (with $k_{max} = \infty$) near the resonant frequency $\omega = |\omega_0|$ [30]. A magnetized plasma has also a pole at $\omega = 0$, but for simplicity here we consider a material response with a single pole. Other magneto-optical materials such as bismuth-iron garnet may be characterized by related material dispersions [31]. The parameter $\omega_e$ defines the strength of the resonance and must have the same sign as $\omega_0$. The objective is to study the edge modes supported by an interface of the nonreciprocal material and a Drude plasma with permittivity $\varepsilon_{\text{Drude}} = 1 - \omega_p^2 / \omega^2$, being $\omega_p$ the plasma frequency.

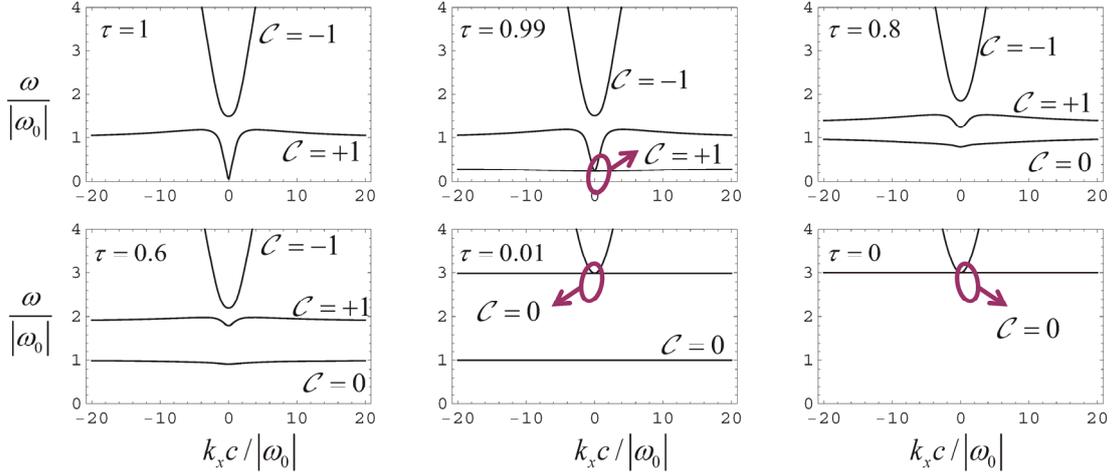

Fig. 1. (Color online) Topological transition between a gyrotropic material with $\omega_e = 0.5\omega_0 > 0$ and $k_{\max} = 10|\omega_0|/c$ ($\tau = 1^-$) and an electric plasma with $\omega_p = 3.0|\omega_0|$ ($\tau = 0^+$). The insets indicate the Chern numbers associated with the relevant photonic bands.

Similar to Ref. [24], it is convenient to introduce an interpolated material response $\varepsilon_\tau(\omega, \mathbf{k}) = \mathbf{1} + (1-\tau)[\varepsilon_{\text{Drude}}(\omega) - \mathbf{1}] + \tau[\varepsilon_M(\omega, \mathbf{k}) - \mathbf{1}]$ where the parameter $\tau$ is such that $0 \leq \tau \leq 1$ and $\varepsilon_M(\omega, \mathbf{k})$ represents the material response determined by Eq. (14). Hence, $\varepsilon_\tau(\omega, \mathbf{k})$ determines a continuous transformation between the Drude plasma ($\tau = 0^+$) and the gyrotropic material ($\tau = 1^-$). Evidently, $\varepsilon_\tau(\omega, \mathbf{k})$ can be decomposed as in Eq. (4) with $\varepsilon_{R,\tau}(\omega) = 1 - (1-\tau)\omega_p^2 / \omega^2$ and $\chi_{\text{NR},\tau}(\omega)$ defined as in Eq. (7) with:

$$\chi_{11} = \chi_{22} = \tau \frac{\omega_0 \omega_e}{\omega_0^2 - \omega^2}, \qquad \chi_{12} = -\chi_{21} = \tau \frac{-i\omega_e \omega}{\omega_0^2 - \omega^2}. \qquad (15)$$

Figure 1 shows the band structure of the TM-polarized modes as the parameter $\tau$ varies continuously from $\tau = 1$ to $\tau = 0$. The insets of the plots indicate the Chern numbers associated with the pertinent photonic bands, which are determined as detailed in Sect. II. As seen, the two materials ($\tau = 1^-$ and $\tau = 0^+$) have a common bandgap and the Chern number associated with the photonic bands above the bandgap is different for each material. As $\tau$ varies continuously from $1^- \to 0^+$, one of the low frequency photonic bands migrates across the bandgap, and this leads to the topological transition. Hence, the two materials ($\tau = 0^+$ and $\tau = 1^-$) are topologically distinct. As already discussed in Refs. [24, 27], to tell if two materials are topologically equivalent or not, one needs to ensure that they belong to the same vector space. Since the relevant vector space is determined by the poles of the material response [24], a convenient way to merge the poles of the materials and ensure that they define the same space is to consider the interpolated material response and the limits $\tau = 0^+$ and $\tau = 1^-$. The topological classification is done using the materials with $\tau = 0^+$ and $\tau = 1^-$.

Using the formalism of Sect. III, we computed the dispersion of the edge modes supported by an interface ($y = 0$) of a medium with $\tau = 0^+$ (Drude plasma in the region $y < 0$) and a medium with $\tau = 1^-$ (gyrotropic material with spatial cut-off in the region $y > 0$) and for different values of $\omega_p$. Consistent with the previous example, in the following the considered materials are topologically distinct and the difference between the Chern numbers associated with the photonic bands below the band gap is $\delta \mathcal{C} = \mathcal{C}_M - \mathcal{C}_{\text{Drude}} = +1$. Hence, the bulk-edge correspondence, if valid, implies that there should exist a single topologically protect gapless edge state in the common gap.

In the first set of examples (Fig. 2), the value of $\omega_p$ is comparable to or smaller than the frequency that determines the upper bandgap edge of the gyrotropic material. As seen in Fig. 2, consistent with the bulk-edge correspondence principle, there is always a topologically protected edge state that spans the entire common gap. The edge state is a backward wave.

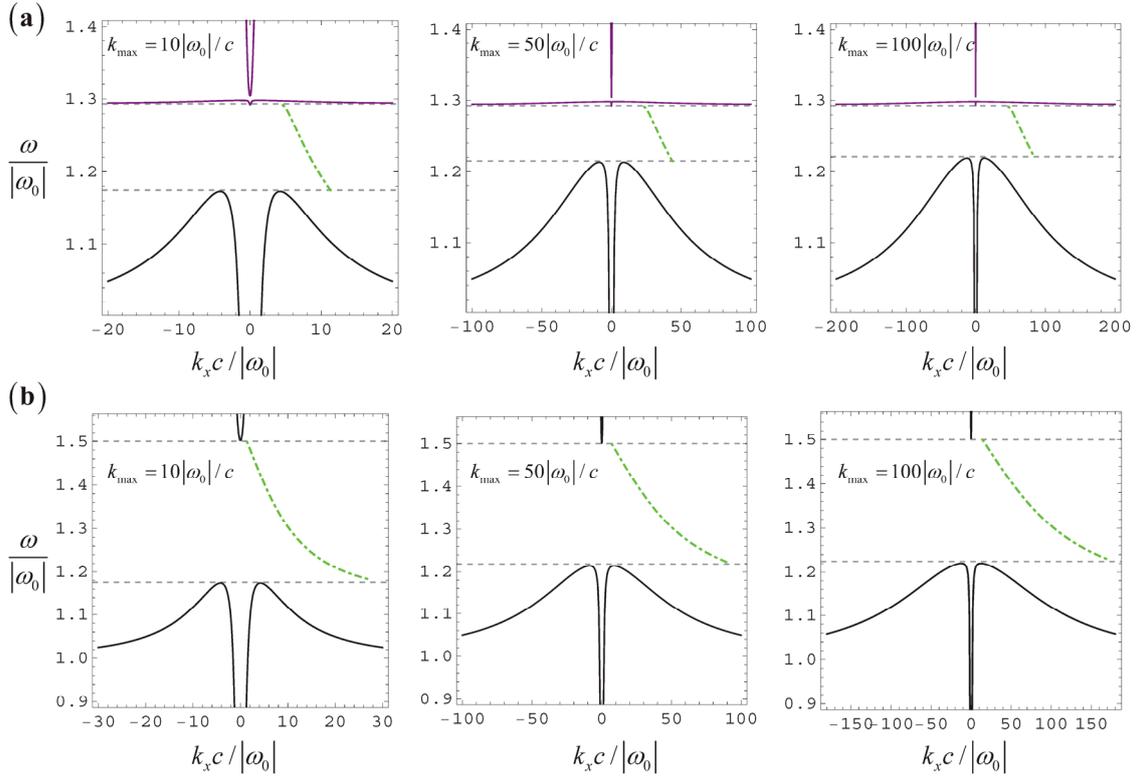

Fig. 2. (Color online) Impact of the high-frequency spatial cut-off $k_{max}$ on the edge modes. Band diagram $\omega$ vs. $k_x$ (green dot-dashed line) for a gyrotropic material with $\omega_e = 0.5\omega_0 > 0$ and an electric plasma with **(a)** $\omega_p = 1.3\omega_0$ and **(b)** $\omega_p = 1.6\omega_0$. The black solid lines (purple solid lines) represent the dispersion of the modes in the bulk gyrotropic material (bulk plasma) with the spatial cut-off. In the limit $k_{max} \to \infty$ the dispersion branch moves to infinity and there are no edge modes. The dashed horizontal gray lines delimit the common bandgap.

Interestingly, as the spatial cut-off is increased, the dispersion of the edge states is shifted to larger spatial frequencies, so that the mode becomes more confined to the interface. In the limit $k_{max} \to \infty$ the edge mode dispersion moves to $k_x \to \infty$ and hence in the limit of a local response ($k_{max} = \infty$) the edge mode disappears and the bulk-edge correspondence does not work. This is one of the main points of the article: the bulk edge correspondence is valid only when the spatial cut-off is properly considered in the calculation of the edge modes. Thus, in the same manner as the spatial cut-off is crucial to obtain integer Chern numbers [24], it must also be considered in the calculation of the edge modes.

For larger values of $\omega_p$, in the range $1.8|\omega_0| < \omega_p < 2.2|\omega_0|$, the dispersion of the one-way edge states has the exotic behavior illustrated in the left and middle panels of Fig. 3a, such that for some frequency in the bandgap the edge states dispersion approaches $k_x \to -\infty$ and then reappears at $k_x \to +\infty$. The branches are thus connected at $k_x = \infty$ (north-pole of the Riemann sphere [24]), so that with a finite spatial cut-off $k_{max}$ the edge states remain gapless and the bulk-edge correspondence is valid. Yet, in the limit of a local response $k_{max} = \infty$ the upper branch of the edges states dispersion diverges to $k_x \to +\infty$, similar to the examples of Fig. 2. In contrast, when $\omega_p > 2.2|\omega_0|$ (see Fig. 3b) the edge states dispersion remains gapless in the local limit, and in this situation the dispersion of the modes with finite $k_{max}$ converges uniformly to the dispersion of the edge states predicted by the local model [Eq. (13)] as $k_{max} \to \infty$. Furthermore, in this case the edge modes are forward waves propagating along the negative $x$-axis. The propagation of unidirectional edge modes in the context of a local material response has been thoroughly studied by other authors [41-44].

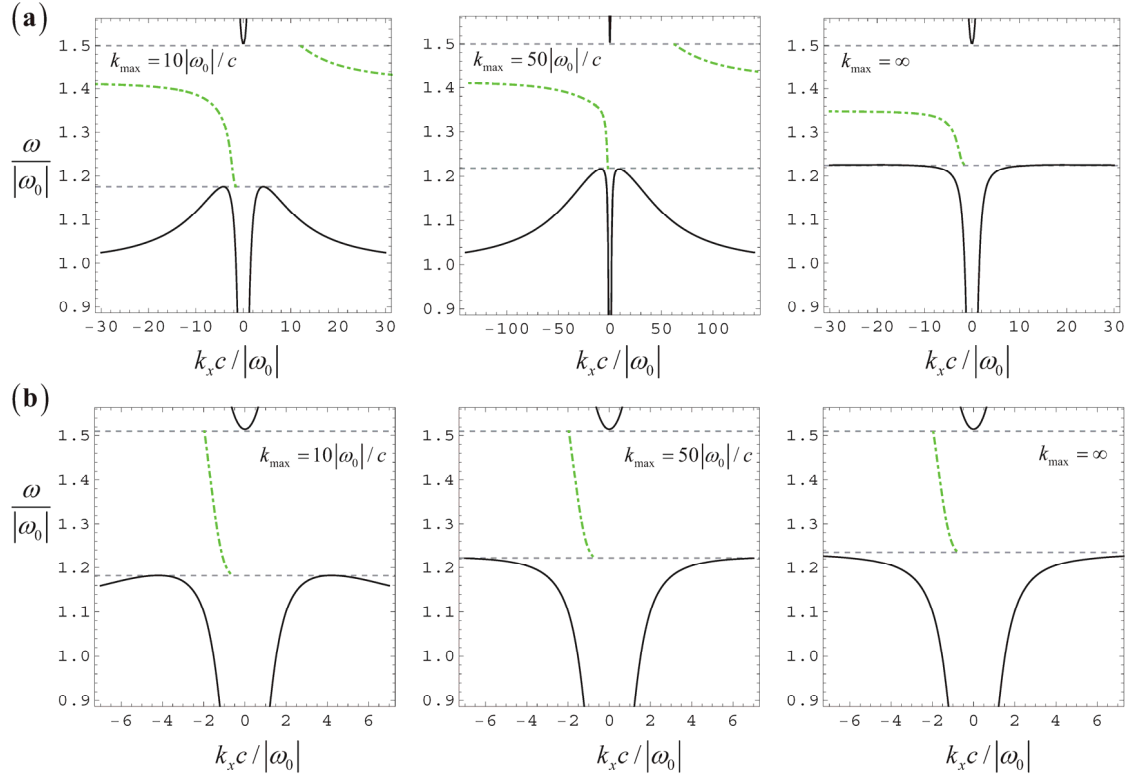

Fig. 3. (Color online) Similar to Fig. 2 but for an electric plasma with **(a)** $\omega_p = 2.0\omega_0$ and **(b)** $\omega_p = 3.0\omega_0$. In panel (a) the two branches of the edge modes dispersion (green curves) are joined at $k_x = \infty$ (the north-pole) in case of a finite $k_{max}$.

The previous results raise the question if topological methods are really useful in the continuum case. Indeed, realistic materials are usually modeled by local constitutive relations (with permittivity independent of the wave vector), and hence the introduction of a spatial cut-off may appear a bit artificial, especially because – as illustrated by the previous examples – the bulk-edge correspondence principle only works when the spatial cut-off is explicitly considered in the edge states calculation. It is worth noting that the *actual* physical response of realistic materials is necessarily characterized by some cut-

-17-

off. For example, for crystalline materials the cut-off may be estimated as $k_{max} \sim 1/a$, being $a$ the lattice constant. Nevertheless $k_{max}$ is typically several orders of magnitude larger than the spatial frequencies that determine the wave phenomena of interest and hence for most purposes conventional media may be regarded as a continuum.

Even though the continuum models of realistic materials are local, it is possible to imitate the spatial cut-off in Eq. (14) by inserting a small air gap with thickness $d$ in between the two materials, as already suggested in our previous work [24]. The logic is that in the $k_{max} \to \infty$ limit the response of the gyrotropic material [Eq. (14)] approaches the vacuum response ($\varepsilon_M(\omega, \mathbf{k}) \to 1$). Hence, we propose here that the spatial cut-off $k_{max}$ can be mimicked by an air gap with thickness $d = 1/k_{max}$ (see Fig. 4). Note that for large $k_x$ the electromagnetic coupling across the gap is determined by a factor of the type $e^{-|k_x|d} = e^{-|k_x|/k_{max}}$.

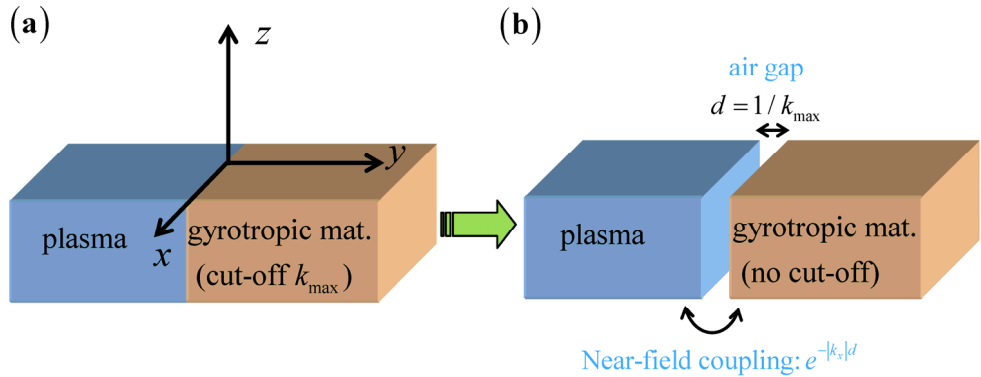

Fig. 4. (Color online) The effect of the spatial cut-off $k_{max}$ in the gyrotropic material response (panel $a$) may be mimicked with an air gap with thickness $d = 1/k_{max}$ (panel $b$).

Interestingly, as illustrated in Fig. 5, the proposed equivalence works fairly well, and in some spectral range the edge modes dispersion calculated with a finite $k_{max}$ is nearly coincident with what is obtained without any cut-off and with an air gap [Eq. (14)]. Hence, this indicates that it is possible to make sense of topological methods in photonic continua, provided the relevant spatial cut-offs are imitated with an air gap. This is the second main point of the article. Even though the correspondence $d \leftrightarrow 1/k_{max}$ gives quite satisfactory results, it should be mentioned that in the example of Fig. 5a the edge modes dispersion does not span the entire gap. In this regard the correspondence between the spatial cut-off and the vacuum gap is not perfect. Yet, the response of the two systems is at least qualitatively rather similar, and consistent with the bulk-edge correspondence principle there is a single edge mode in the common bandgap. Note that without the air gap there are no edge modes in the example of Fig. 5a.

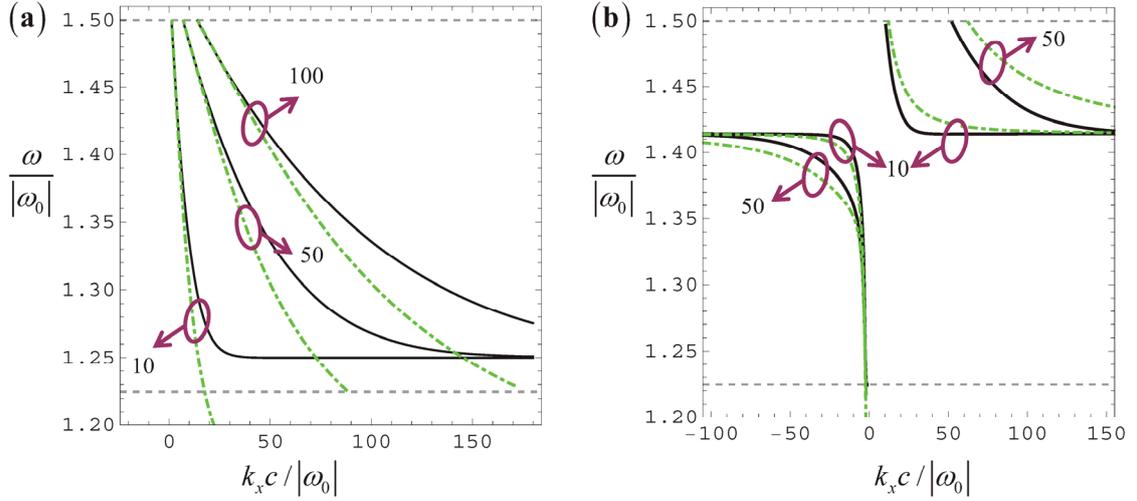

Fig. 5. (Color online) Edge modes for an interface between a gyrotropic material with a spatial cut-off and an electric plasma (green dot-dashed lines) superimposed on the modes supported by the same materials with no cut-off separated by an air layer with $d = 1/k_{max}$ (black solid lines). The insets give the value of $k_{max} c / |\omega_0|$. (a) $\omega_p = 1.6\omega_0$. (b) $\omega_p = 2.0\omega_0$. The dashed horizontal gray lines delimit the common bandgap.

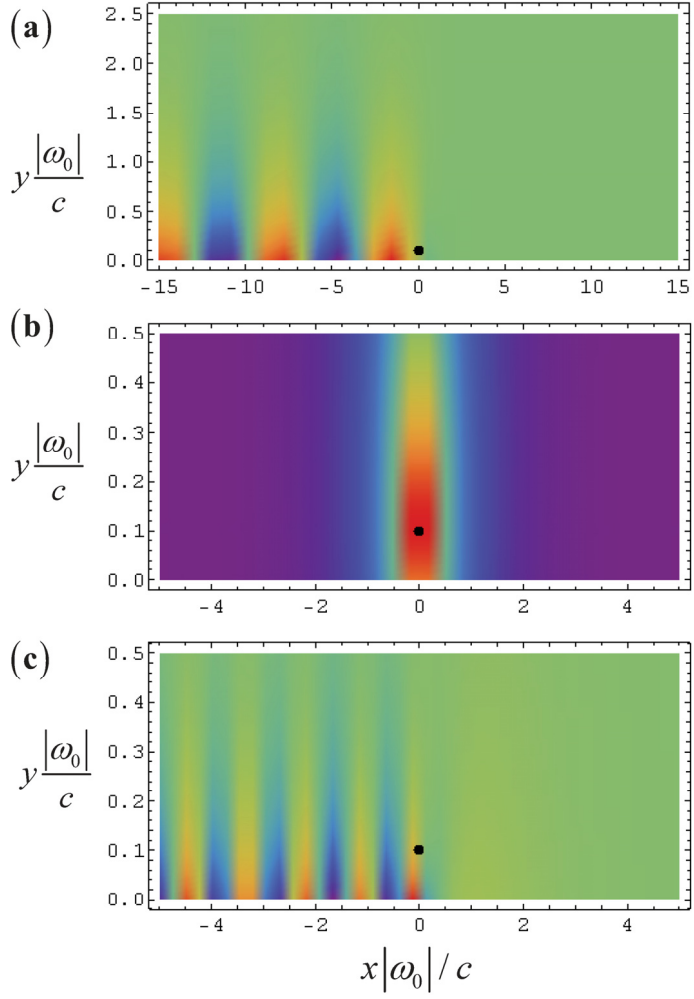

Fig. 6. (Color online) Time snapshot of the magnetic field emitted by a line source (black dot in the figures) embedded in a gyrotropic material with $\omega_e = 0.5\omega_0 > 0$ (region $y > 0$). The oscillation frequency is $\omega = 1.4\omega_0$. **(a)** The region $y < 0$ is a perfect electric conductor. **(b)** The region $y < 0$ is a Drude plasma with $\omega_p = 1.6\omega_0$. **(c)** Similar to **b)** but an air gap with thickness $d = 0.1c/|\omega_0|$ is inserted in between the two materials to mimic a spatial cut-off.

To further illustrate the importance of considering the air gap to mimic the spatial cut-off Fig. 6 represents a time snapshot of the magnetic field ($H_z$) emitted by a magnetic line source placed at a height $h = 0.1c/|\omega_0|$ above the interface. In this example, the materials response is assumed local ($k_{max} = \infty$). The line source is embedded in the gyrotropic material and the oscillation frequency lies within the common bandgap. The radiated fields are obtained with the formalism of Appendix A. The plot depicts $H_z$ only in the gyrotropic material (region $y > 0$). In the first case (Fig. 6a), the region $y < 0$ is taken as a perfect electric conductor ($\omega_p = \infty$). In this case it is irrelevant to consider the air gap because as previously discussed for $\omega_p > 2.2|\omega_0|$ the dispersion of the edge modes converges uniformly to the local theory result as $k_{max} \to \infty$. Since none of the materials has bulk states at the frequency of oscillation the only possible radiation channel is determined by the edge modes. Consistent with this property, one sees in Fig. 6a that the line source excites a unidirectional topologically protected edge state that propagates along the $-x$ direction. When the bottom region is replaced by a Drude plasma with $\omega_p = 1.6\omega_0$ (Fig. 6b) the line source is unable to excite edge modes because without a spatial cut-off the material interface does not support edge states. Importantly, if the spatial cut-off is imitated with an air gap it becomes possible to excite an edge mode (Fig. 6c), consistent with the bulk-edge correspondence principle (see also Figs. 2b and 5a).

To further explore the use of topological methods in photonic systems, next it is shown that it is possible to pair electromagnetic continua that give rise to a Chern number difference greater than one, $|\delta \mathcal{C}| > 1$ (high order Chern number). To demonstrate this possibility, we consider two distinct gyrotropic materials, with a response consistent with Eq. (14), and parameters $\omega_{01} = -\omega_{02}$ and $\omega_{e1} = -\omega_{e2}$. Note that flipping the sign of $\omega_0$ corresponds to flipping the sign of the biasing magnetic field. Thus, the two material responses may be obtained in practice using a single material and symmetric biasing fields. It is interesting to note that this system is reminiscent of the Haldane model [45], in the sense that the average nonreciprocal response of the system (the biasing magnetic field averaged in space) is zero. A configuration related to ours but for a two-dimensional electron gas was studied in Ref. [26]. The emergence of topological effects in photonic crystals with a spatially varying biasing field was also discussed in Ref. [13], and the experimental observation of high order Chern numbers was reported in Ref. [12].

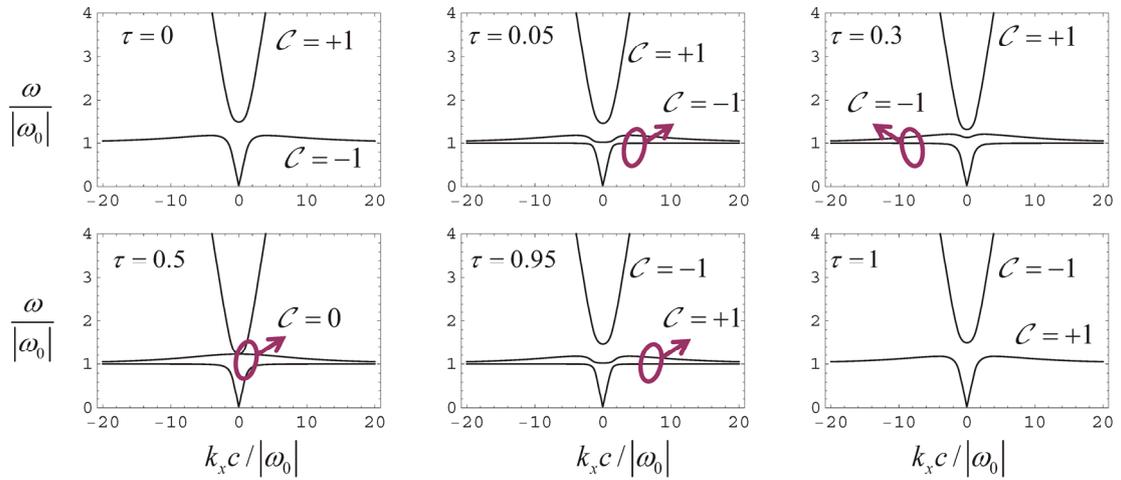

Fig. 7. (Color online) Topological transition between a gyrotropic material with $\omega_{e1} = 0.5\omega_{01}$ ($\tau = 1$) and a gyrotropic material with $\omega_{e2} = 0.5\omega_{02}$ ($\tau = 0$) with $\omega_{01} = -\omega_{02} \equiv \omega_0 > 0$. The spatial cut-off is $k_{max} = 10|\omega_0|/c$. The insets indicate the Chern numbers associated with the relevant photonic bands.

The band diagrams of the interpolated material response for different values of $\tau$ are represented in Fig. 7. As seen, the initial and final material responses ($\tau = 0$ and $\tau = 1$) give rise to identical band diagrams, but with symmetric Chern numbers. Thus, the Chern number difference is $\delta \mathcal{C} = \mathcal{C}_{\tau=1^-} - \mathcal{C}_{\tau=0^+} = 2$. The topological transition occurs exactly for $\tau = 0.5$, when the bandgap closes and reopens, and the Chern numbers are exchanged by the different bands.

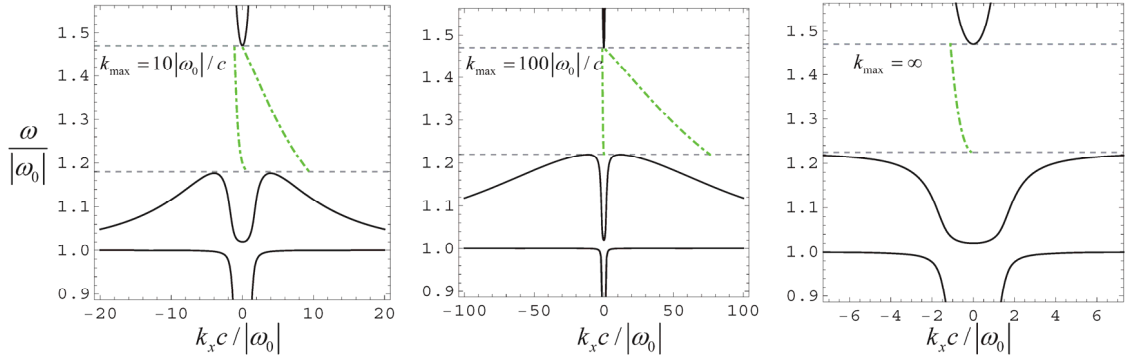

Fig. 8. (Color online) Edge modes at an interface between two topologically distinct gyrotropic materials with the same parameters as in Fig. 7 for $k_{max} = 10|\omega_0|/c$, $k_{max} = 100|\omega_0|/c$, and $k_{max} = \infty$.

Figure 8 shows the dispersion of the edge modes supported by an interface of the two topologically distinct materials. Consistent with the bulk-edge correspondence principle, when the spatial cut-off is taken into account (left and middle panels) there are two gapless unidirectional edge states. However, similar to the first examples in this section, in the $k_{max} \to \infty$ limit the dispersion of one of the modes migrates to infinity. Hence, for local material responses ($k_{max} = \infty$) the structure supports a single edge state (rightmost panel in Fig. 8) and the bulk-edge correspondence principle fails. However, by mimicking the spatial cut-off with an air gap the bulk-edge correspondence is recovered, as illustrated in Fig. 9. Indeed, with an air gap the system supports precisely two unidirectional edge modes. The dispersion of the mode that migrates to infinity is particularly well imitated by the system with the air gap for relatively small values of $k_x$ ($|k_x| < 0.5 k_{max}$), which from a practical point of view is typically the most interesting situation. Indeed, modes with small $k_x$ tend to be less affected by realistic material loss. This property – which is also manifest in Fig. 5 – is justified by the fact that an air gap introduces a spatial cut-off law with an exponential decay ($e^{-|k_x|/k_{max}}$), whereas the model

-23-

in Eq. (14) leads to an algebraic decay. Thus, in general it is expected that the air gap mimics better the spatial cut-off for values such that $|k_x| \sim k_{max}$, whereas for $|k_x| \gg k_{max}$ the agreement may be less satisfactory.

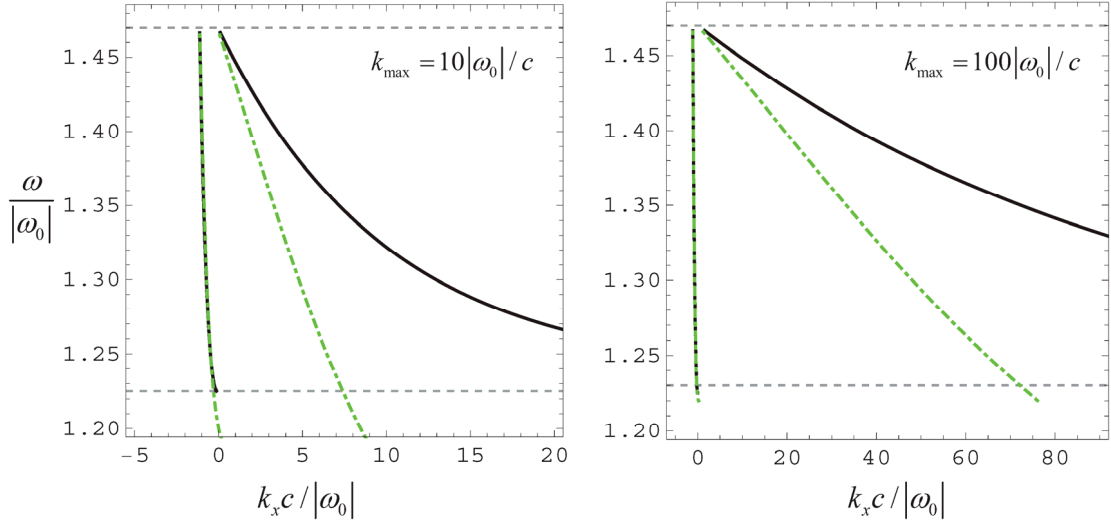

Fig. 9. (Color online) Edge modes at an interface between two topologically distinct gyrotropic materials with a spatial cut-off frequency (green dot-dashed lines) superimposed on the modes supported by the same materials with no spatial cut-off separated by an air gap with thickness $d = 1/k_{max}$ (black solid lines).

## V. Conclusion

It was highlighted that the use of topological methods in electromagnetic continua, and in particular the validity of the bulk-edge correspondence principle, in general require that the (nonreciprocal part) of the material response has a spatial cut-off. It was demonstrated that when the spatial cut-off is explicitly considered and the nonlocal response is properly taken into account the bulk-edge correspondence is valid. It was shown that in practice the spatial cut-off can be implemented by inserting a tiny air gap in between the materials with thickness inversely proportional to the cut-off. Thus, the proposed ideas open new inroads in topological photonics and provide a roadmap to explore the topological properties of systems formed by electromagnetic continua with no intrinsic periodicity.

## *Appendix A: Fields radiated by the line source*

Here, the fields emitted by a magnetic line source embedded in a gyrotropic material are calculated. The line source is located at $(x_0, y_0) = (0, h)$ and is infinitely extended along the z-direction. The region $y > 0$ is occupied by the gyrotropic material, whereas the region $y < -d$ is filled with a Drude plasma. Here, $d$ denotes the thickness of an air gap defined by $-d < y < 0$.

For a magnetic line source excitation, the Maxwell's equations read:

$$\nabla \times \mathbf{E} = i\omega\mu_0 \mathbf{H} - \mathbf{j}_m, \qquad \nabla \times \mathbf{H} = -i\omega\varepsilon_0 \boldsymbol{\varepsilon} \cdot \mathbf{E}, \qquad (A1)$$

where $\mathbf{j}_m = I_m \delta(x)\delta(y - y_0)\hat{\mathbf{z}}$ is the (fictitious) magnetic current density and $I_m$ is the equivalent magnetic current. Looking for a solution of the form $\mathbf{H} = H_z(x,y)\hat{\mathbf{z}}$ it is found that for $y > 0$:

$$\nabla^2 H_z + \left(\frac{\omega}{c}\right)^2 \varepsilon_{ef} H_z = -A_0 \delta(x)\delta(y - y_0), \qquad A_0 = i\omega\varepsilon_0 \varepsilon_{ef} I_m, \qquad (A2)$$

where $\varepsilon_{ef} = (\varepsilon_{11}^2 + \varepsilon_{12}^2)/\varepsilon_{11}$ is the equivalent permittivity of the gyrotropic material. Thus, in a bulk medium the emitted field is (when the whole space is filled with the gyrotropic material):

$$H_z^{inc} = \frac{A_0}{-4i} H_0^{(1)}\left(\frac{\omega}{c}\sqrt{\varepsilon_{ef}}\rho\right) = A_0 \frac{1}{2\pi} \int \frac{1}{2\gamma_M} e^{-\gamma_M |y - y_0|} e^{ik_x x} dk_x, \qquad (A3)$$

where $H_0^{(1)}$ is the Hankel function of first kind and order zero, $\rho = \sqrt{x^2 + (y-y_0)^2}$ and $\gamma_M = \sqrt{k_x^2 - (\omega/c)^2 \varepsilon_{ef}}$. The integral form of the emitted field can be easily modified to take into account the presence of an interface at $y=0$:

$$H_z = A_0 \frac{1}{2\pi} \int \frac{1}{2\gamma_M} \left( e^{-\gamma_M |y-y_0|} + R e^{-\gamma_M (y+y_0)} \right) e^{ik_x x} dk_x, \tag{A4}$$

where $R = R(\omega, k_x)$ represents the (magnetic field) reflection coefficient for TM-polarized waves. For the geometry of interest, it is straightforward to show that:

$$R = \frac{\left( \frac{\varepsilon_{11}}{\varepsilon_{11}^2 + \varepsilon_{12}^2} \gamma_M + \frac{\varepsilon_{12}}{\varepsilon_{11}^2 + \varepsilon_{12}^2} ik_x \right)\left( 1 + \frac{\gamma_D}{\varepsilon_D} \frac{\varepsilon_{air}}{\gamma_{air}} \tanh(\gamma_{air} d) \right) - \left( \frac{\gamma_D}{\varepsilon_D} + \frac{\gamma_{air}}{\varepsilon_{air}} \tanh(\gamma_{air} d) \right)}{\left( \frac{\varepsilon_{11}}{\varepsilon_{11}^2 + \varepsilon_{12}^2} \gamma_M - \frac{\varepsilon_{12}}{\varepsilon_{11}^2 + \varepsilon_{12}^2} ik_x \right)\left( 1 + \frac{\gamma_D}{\varepsilon_D} \frac{\varepsilon_{air}}{\gamma_{air}} \tanh(\gamma_{air} d) \right) + \left( \frac{\gamma_D}{\varepsilon_D} + \frac{\gamma_{air}}{\varepsilon_{air}} \tanh(\gamma_{air} d) \right)}. \tag{A5}$$

In the above, $\varepsilon_{11}, \varepsilon_{12}$ determine the response of the gyrotropic material, $\varepsilon_D$ is the permittivity of the metal (described by a Drude model), $\gamma_D = \sqrt{k_x^2 - \varepsilon_D (\omega/c)^2}$, and $\varepsilon_{air} = 1$ and $\gamma_{air} = \sqrt{k_x^2 - \varepsilon_{air}(\omega/c)^2}$ determine the parameters of the air gap.

**Acknowledgements:** This work was funded by Fundação para a Ciência e a Tecnologia under project PTDC/EEI-TEL/4543/2014.